\documentclass[final,1p,times]{elsarticle} 
\usepackage{graphicx}
\usepackage{amssymb} 
\usepackage{amsthm} 
\usepackage{lineno}

\newcommand{\gevc}{GeV/$c$}
\newcommand{\pt}{p_{\rm T}}
\newcommand{\deta}{\Delta\eta}
\newcommand{ \mean }[1]{\left\langle #1 \right\rangle}   

\journal{Nuclear Physics A} 

\begin{document}

\begin{frontmatter} 

% Your Title - please insert
\title{Results on flow from the ALICE Collaboration}

%% Single author (and collaboration) - please insert
\author{Sergei A. Voloshin (for the ALICE\fnref{col1} Collaboration)}
\fntext[col1] {A list of members of the ALICE Collaboration and
  acknowledgements can be found at the end of this issue.}
\address{Wayne State University, 666 W. Hancock, Detroit, Michigan
  48201}

%% Multiple authors
%\author[auth2]{Marcus Junius Brutus}
%\address[auth1]{Somewhere, Rome}
%\address[auth2]{Somewhere else, Rome}

\begin{abstract} 
This short overview includes recent results from the ALICE
Collaboration on anisotropic flow of charged and identified particles
in $\sqrt{s_{\rm {NN}}}=2.76$~TeV Pb-Pb collisions. We also discuss
charge dependent and event plane dependent azimuthal correlations that
are important in tests of the chiral magnetic effect, as well as
understanding the dynamics of the system evolution and hadronization
process. Lastly, we present ALICE results obtained with a new
technique, the event shape engineering, which allows to perform a
physical analysis on events with very large or small flow.

\end{abstract} 

\end{frontmatter} % do not change

%% linenumbers are useful for reviewing process
%\linenumbers

%============================================================
\section{Introduction}

Anisotropic flow was one of the most expected results from heavy-ion
program at the LHC. The very first measurements have shown that the
anisotropic flow at LHC is large and in rough agreement with
expectation from the hydrodynamic model. These results have been
intensively discussed at the last Quark Matter
conference~\cite{qm2011, Muller:2012zq} and have been
published~\cite{Aamodt:2010pa,ALICE:2011ab}, but did not reduce the
interest in flow results, as new questions have risen up, in
particular the ones related to the initial geometry
fluctuations. Also, high-energy and high-multiplicity, as well
high-statistics data from LHC allow for unprecedented study of event
anisotropies at high transverse momenta as well as precision
measurements of multiparticle (higher cumulants) correlations.
Identified particle flow, for which the ALICE experiment~\cite{karel}
is suited the best among all LHC experiments, is also highly
interesting as it provides important information for further
understanding of the system evolution (e.g. development of radial
flow) as well as hadronization process.  Below we discuss new ALICE
results related to these questions.  The results on elliptic flow of
${\rm J}/\psi$, $\rm D$-mesons, and muons from heavy flavor decays can
be found in~\cite{jpsi,dmeson,muon}.

In recent years, the charge dependence of the azimuthal correlations,
and in particular in correlations with respect to the reaction plane
has attracted a lot of attention, first as a measurement sensitive to
the Chiral Magnetic Effect (CME)~\cite{Kharzeev:2004ey}, but also due
to its sensitivity to the system hadronization and, more generally, to
the system evolution till the freeze-out, e.g. charge (quark)
production and charge diffusion in the system during its evolution.
ALICE Collaboration measurements~\cite{Abelev:2012pa} of the charge
dependent correlations are consistent with the (mostly qualitative)
expectations from the CME. At the same time the same data would be
consistent with the effect of the local charge conservation in
combination with the large elliptic flow. ALICE is performing
measurements which should allow to clarify the picture. Some of those
are discussed in these proceedings.

Finally, we review ALICE preliminary results using a new technique, 
the so-called event shape engineering~\cite{Schukraft:2012ah}.
This approach allows to select events of a
particular shape, either more asymmetric or more ``round'', the
direction to be very promising in many respects. In this
study we address such questions as the role of event shape
fluctuations on flow at large transverse momenta. The results are
in agreement with results obtained with other methods, in particular
comparing 4-particle cumulant results with event plane and 2-particle
cumulant measurements.

%========================================================================
\section{Extending the measurements}

%----------------------------------------------------------------
%\subsection{Larger pseudorapidity coverage and higher transverse momenta}

\begin{figure}[htbp]
\begin{center}
\includegraphics[width=0.46\textwidth]{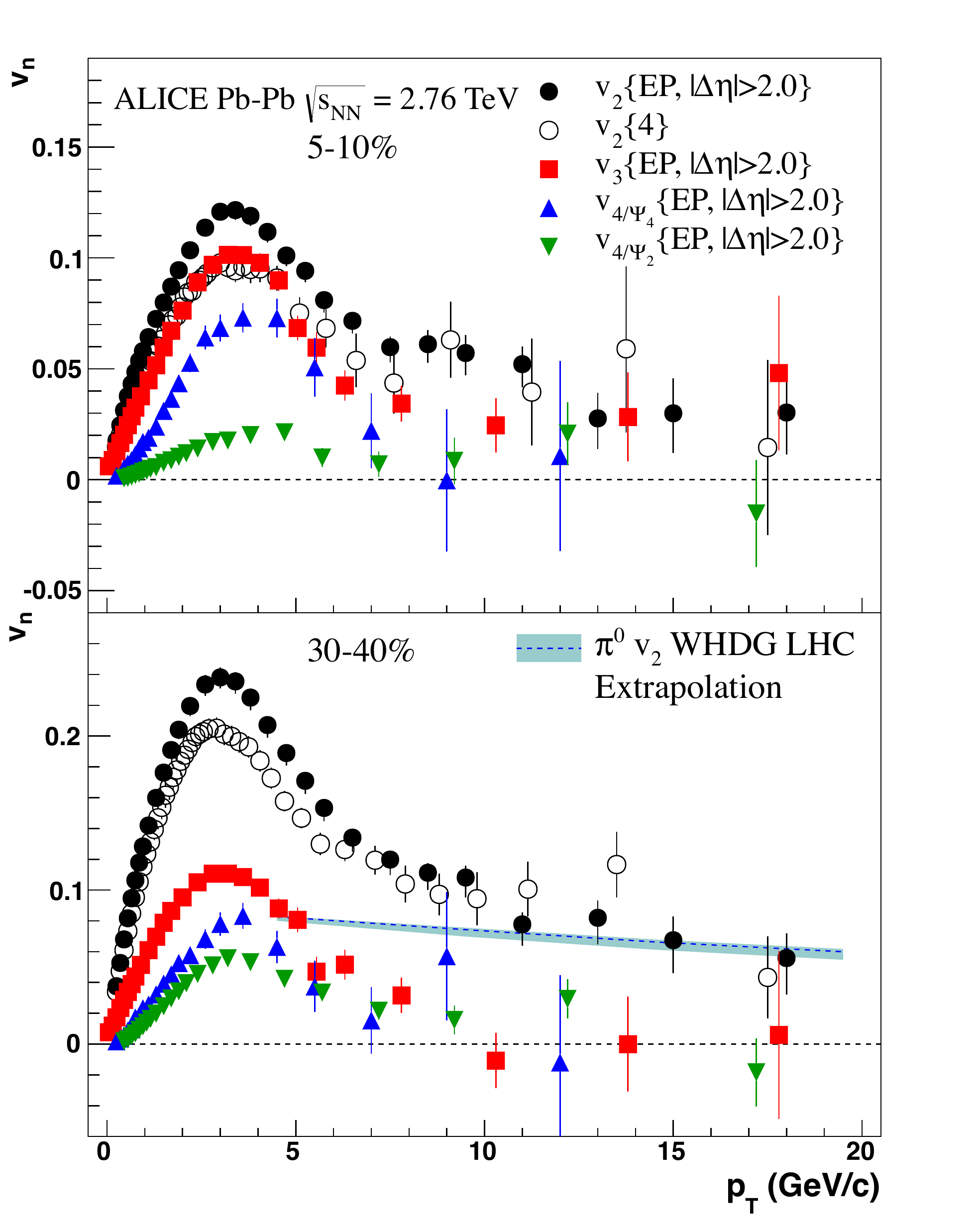}
\includegraphics[width=0.52\textwidth]{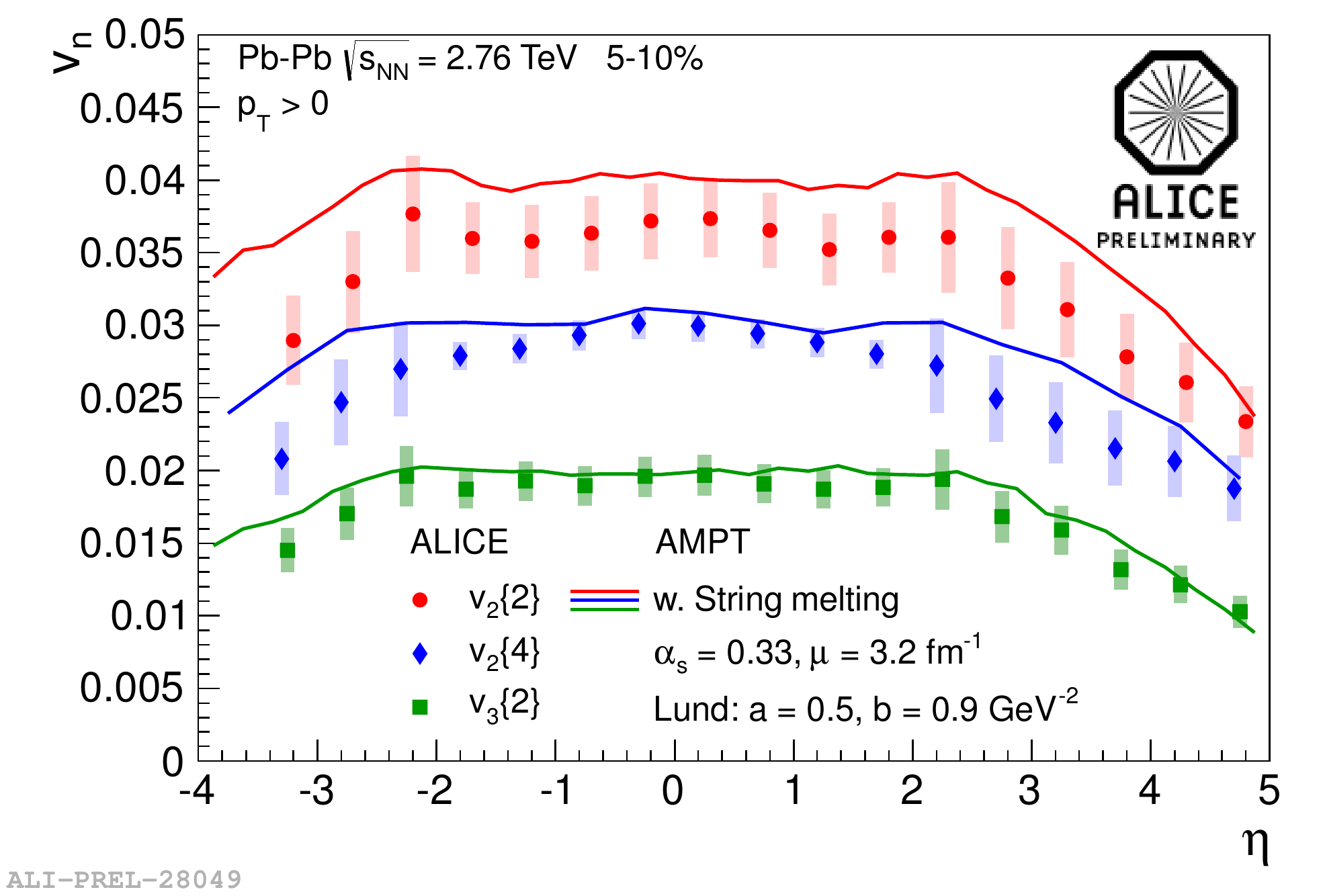}
\end{center}
\caption{(Left panel) $\pt$-differential event
  anisotropies~\cite{Abelev:2012di}. (Right panel) $v_2$ and $v_3$
  pseudorapidity dependence.  }
\label{fig:pteta}
\end{figure}

Recently, ALICE has submitted the paper~\cite{Abelev:2012di} extending
its anisotropic flow measurements up to $\pt\approx 20$~GeV/c.  The
nonflow effects in this analysis has been suppressed by using the
event plane from the VZERO detectors separated from the TPC by two units of
rapidity with an estimate of remaining nonflow included in the
systematic error.  Along with elliptic flow, ALICE reported results on
triangular flow $v_3$ and fourth harmonic anisotropy measured with
respect to both second and fourth order event planes, $v_{4/\Psi_2}$
and $v_{4/\Psi_4}$.  The difference between the two is totally due to
fluctuations in the fourth order harmonic flow and as such provides
important constraints on the physics and origin of the flow
fluctuations. Figure~\ref{fig:pteta} shows the results for the charge
particle anisotropies. Results for pions and protons has been also
measured (see~\cite{Abelev:2012di,noferini}).  Significant nonzero
elliptic flow as well as $v_3$ were found up to the highest transverse
momenta. At $\pt>10$~\gevc~the elliptic flow results are well
described by the WHDG model extrapolation to the LHC
energies~\cite{Horowitz:2011gd} taking into account collision and
radiative energy loss in the expanding medium.

The analysis of the data from the Forward Multiplicity and the Silicon
Pixel Detectors (for details, see~\cite{hansen}) has allowed to extend
the measurements of average (no transverse momentum information)
anisotropy, $v_2$ and $v_3$ up to $\eta=5$, well above the existed so far
measurements. Noteworthy that elliptic flow measurements have been performed
with two- as well as four-particle
cumulant methods, which allow to study flow fluctuations up to very
forward pseudorapidities (see below).  Anisotropic flow has been found
to be almost independent of pseudorapidity in the region $|\eta|<2$,
see Fig.~\ref{fig:pteta} right panel, and decreasing at larger
pseudorapidities. The latter allows to test the limiting
fragmentation (longitudinal scaling) picture, the property of the
spectra showing the scaling with $\eta-y_{beam}$, the observation made
first at lower collision energies.  ALICE results agree well with
such a scaling.  More details of this
analysis can be found in~\cite{hansen}.

%----------------------------------------------------
%\subsection{Identified particle flow}

The shape of the $\pt$-differential anisotropic flow suggests an
existence of several regions in the transverse momentum space with
distinctly different underlying physics. It is now widely accepted
that at $\pt<1$--2 \gevc~the flow pattern is mostly determined by
hydrodynamical flow exhibiting typical ``mass
splitting''~\cite{Huovinen:2001cy}. At large transverse momenta,
$\pt>10$~\gevc, the anisotropy is believed to be defined by the jet
quenching mechanism. One would expect very little particle type
dependence in this region, but unfortunately there exists no 
calculations for that.  The intermediate $\pt$ region is less
understood. It has been observed that in this region 
all baryons have
similar flow which differs from meson flow approximately in a 3:2
ratio. Such a scaling finds a natural explanation in the quark
coalescence picture, the so-called Number of Constituent Quark (NCQ)
scaling~\cite{Voloshin:2002wa}. The exact boundaries, where this
picture is valid, if at all, is not exactly known. As the 
quark anisotropic flow and hadronization via coalescence means the
system being in a deconfined stage, the observation of such a scaling
is very important. 
%Out of all LHC Collaborations, ALICE is the best
%suited for such a study.

\begin{figure}[htbp]
\begin{center}
\includegraphics[width=0.48\textwidth]{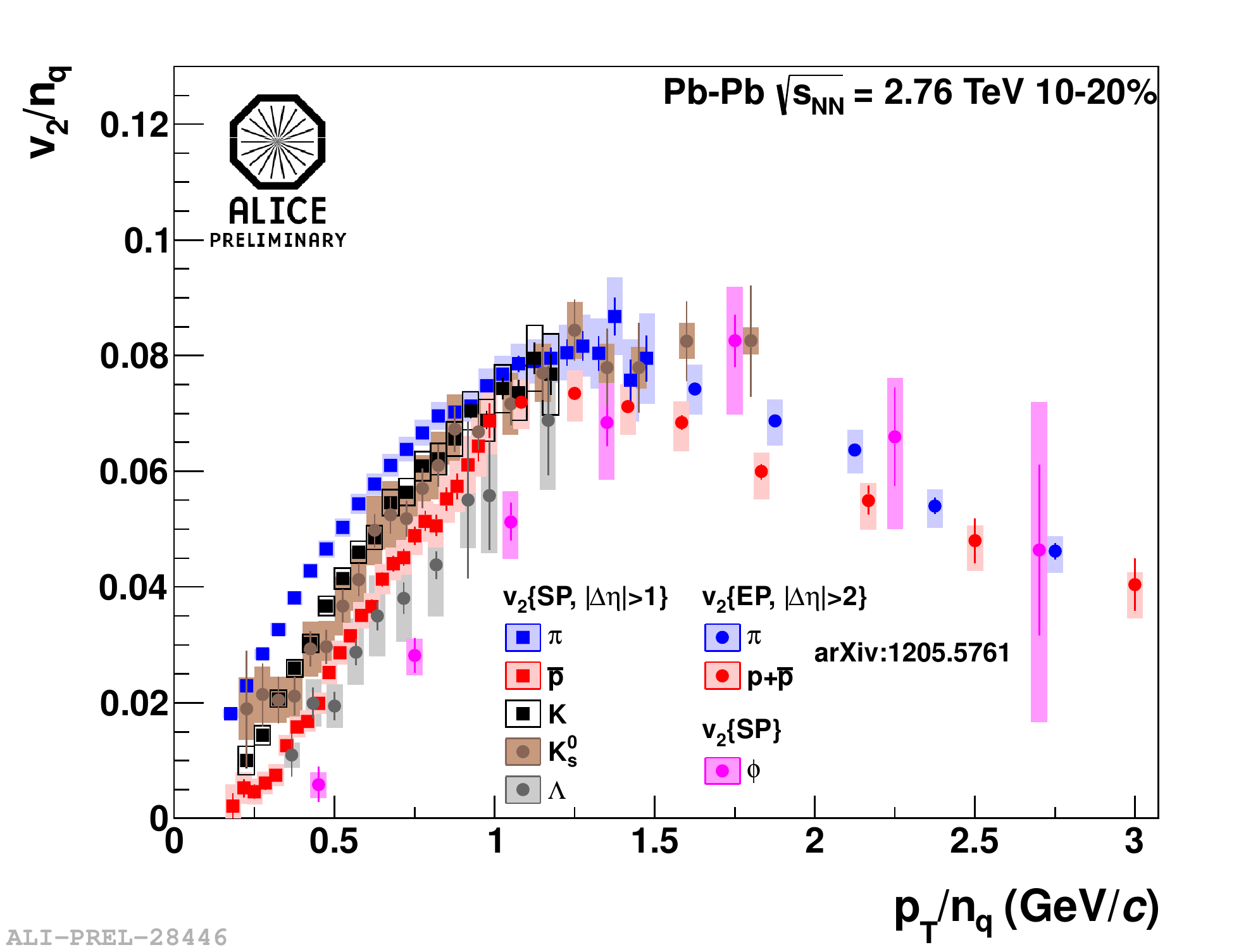}
\includegraphics[width=0.48\textwidth]{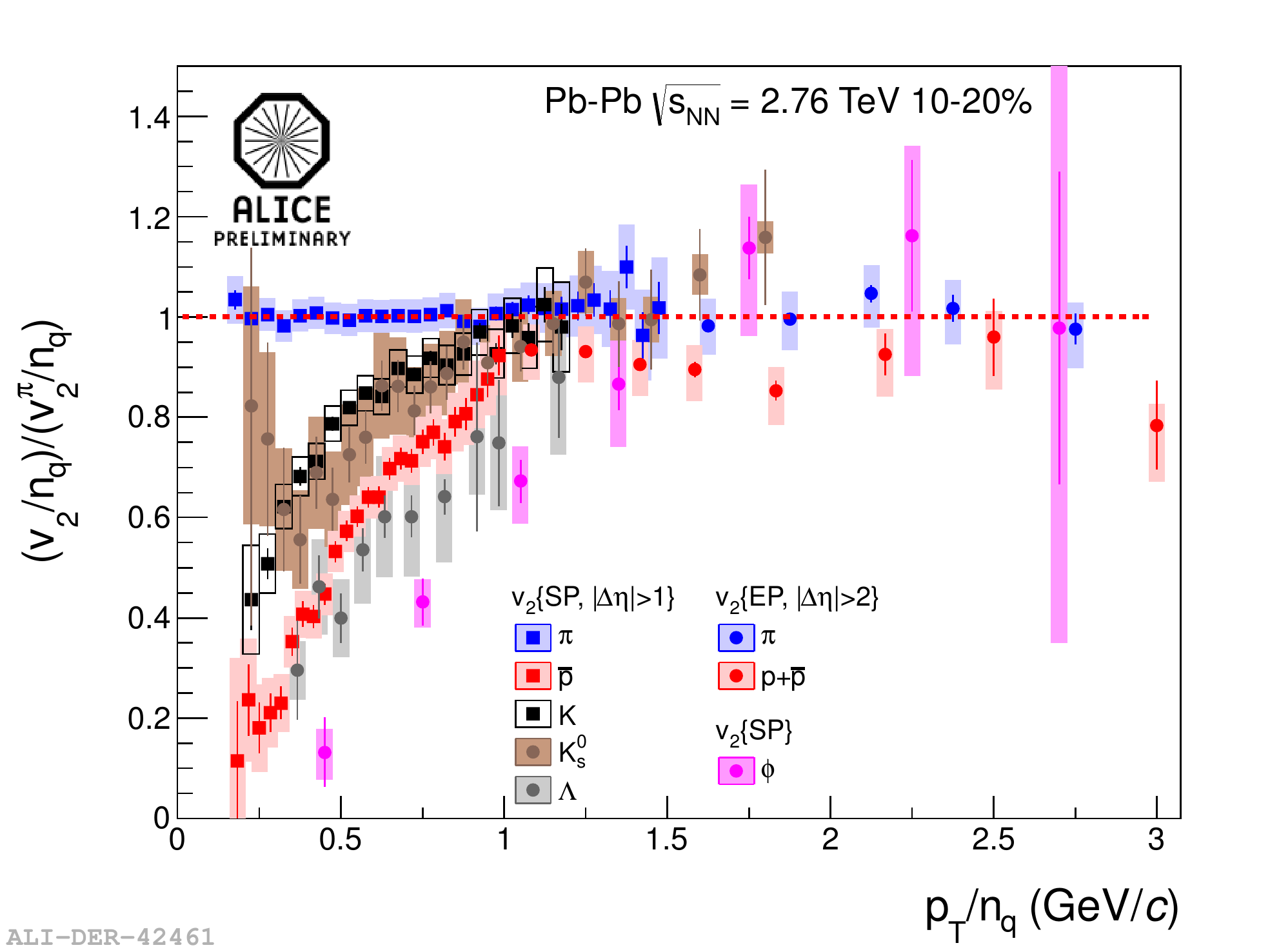}
\end{center}
\vspace*{-4mm}
\caption{(Left panel) Identified particle elliptic flow scaled with the number
  of constituent quarks. (Right panel) the same data divided by the
  polynomial fit to the pion elliptic flow. }
\label{fig:pid}
\end{figure}

Figure~\ref{fig:pid} presents the ALICE results on identified particle
flow, where both, $\pt$~and $v_2$ are scaled by the number of
constituent quarks. Remarkable is the flow of $\phi$-meson, which
exhibits even smaller $v_2$ than expected mass dependence at low
transverse momenta, but scales very much as other mesons at large
$\pt$. One can judge how well the NCQ scaling holds from the right
panel of Fig.~\ref{fig:pid}.  In the range $\pt/n_q >1$~\gevc~the
scaling seems to hold at the level of about 10-15\%.
For a more detailed discussion of the anisotropic flow of identified
particles and comparison to hydrodynamic models, see~\cite{noferini}.

%=====================================================================
\section{Understanding flow fluctuations}

\begin{figure}[htbp]
\begin{center}
\includegraphics[width=0.48\textwidth]{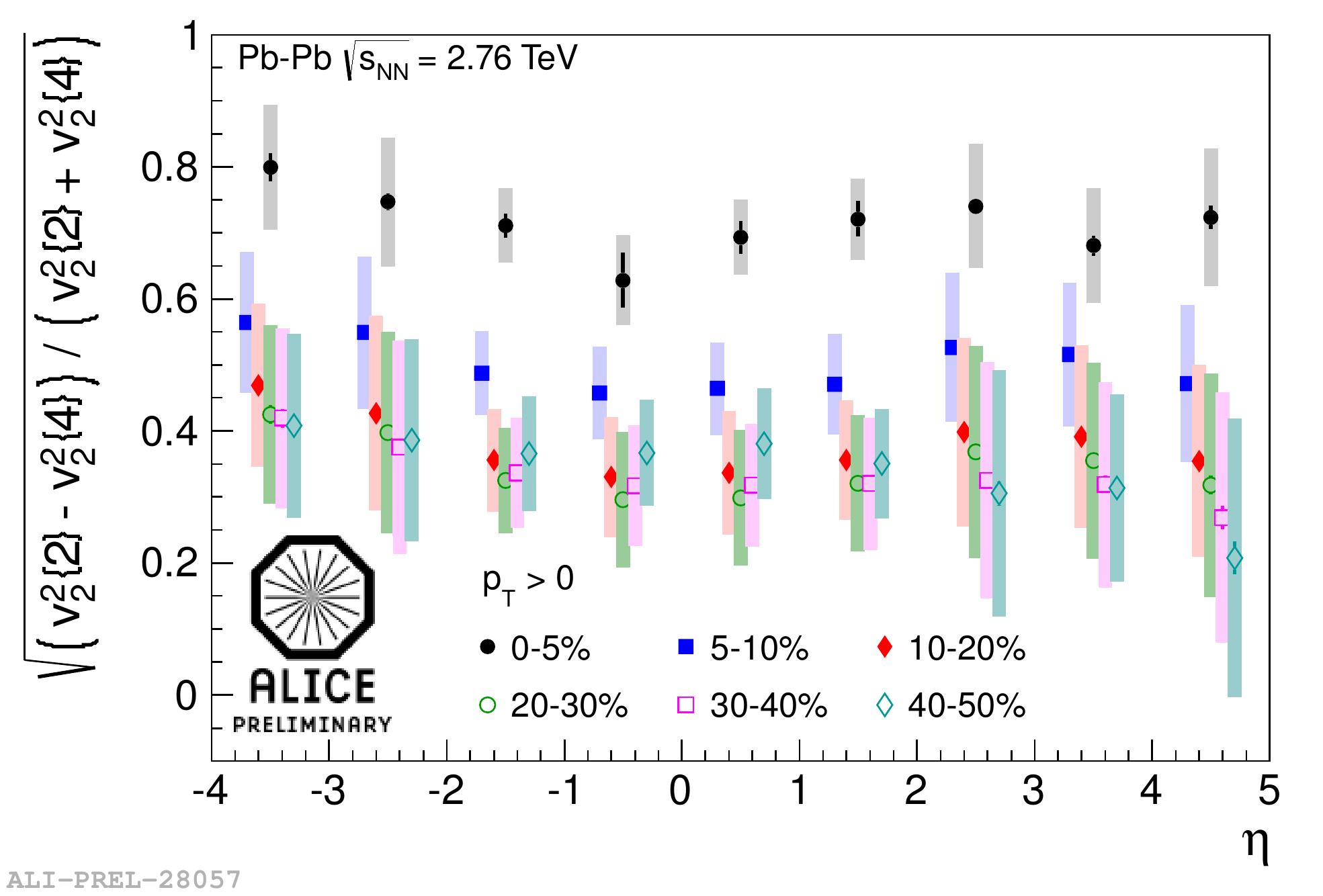}
\includegraphics[width=0.48\textwidth]{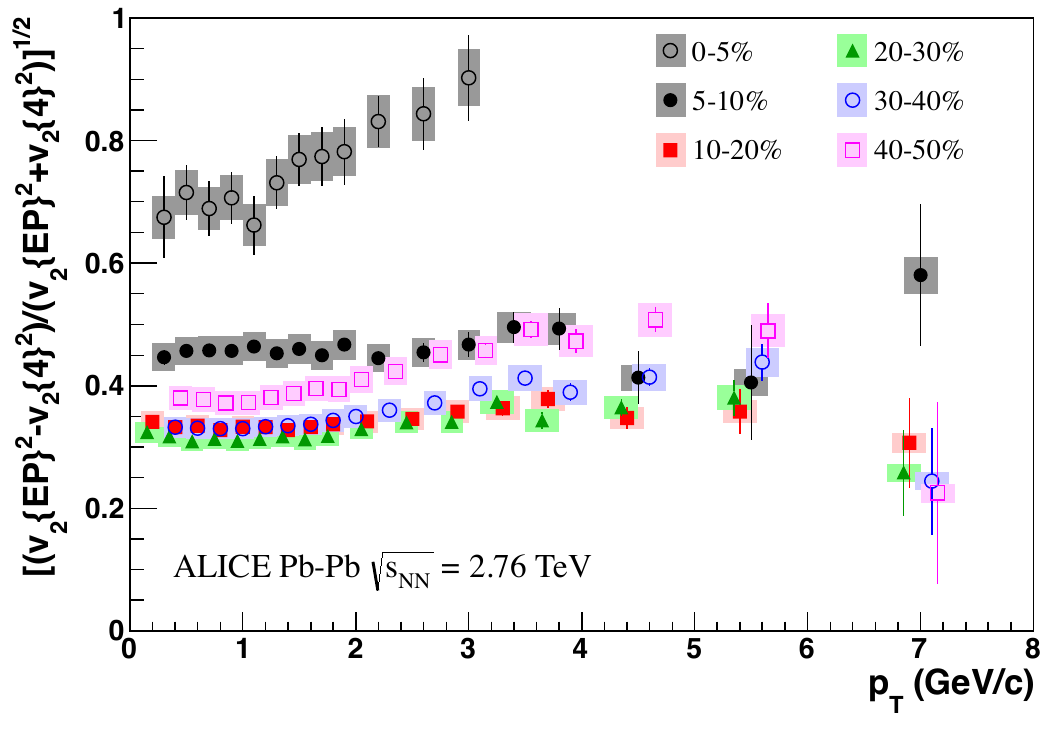}
\end{center}
\vspace*{-4mm}
\caption{Relative flow fluctuation as a function of transverse momentum
  (left panel) and pseudorapidity (right panel) for different
  collision centrality classes.}
\label{fig:ffluc}
\end{figure}

The anisotropic flow measurements performed by ALICE with both two-
and many-particle cumulant methods in wider pseudorapidity regions and
at large transverse momenta allow unprecedented tests of the role of
flow fluctuations.  The flow fluctuations, believed to originate in
the fluctuations in the initial geometry, can be estimated from the
difference in $v_n\{2\}$ and $v_n\{4\}$~\cite{Voloshin:2008dg}.  The
results are presented in Fig.~\ref{fig:ffluc}. Flow
fluctuations are found almost independent on pseudorapidity at all
centralities and very similar at different transverse momenta up to
$\pt\sim~10$~\gevc. The uncertainties are too large at larger
transverse momenta, where the effect of fluctuations might be
small. A similar conclusion on $\pt$ extent of flow
fluctuations can be drawn from the comparison of $v_4$ measured with
respect to the second and fourth order event planes, which become
similar at $\pt>10$~\gevc.

\begin{figure}[htbp]
\begin{center}
\includegraphics[width=0.48\textwidth]{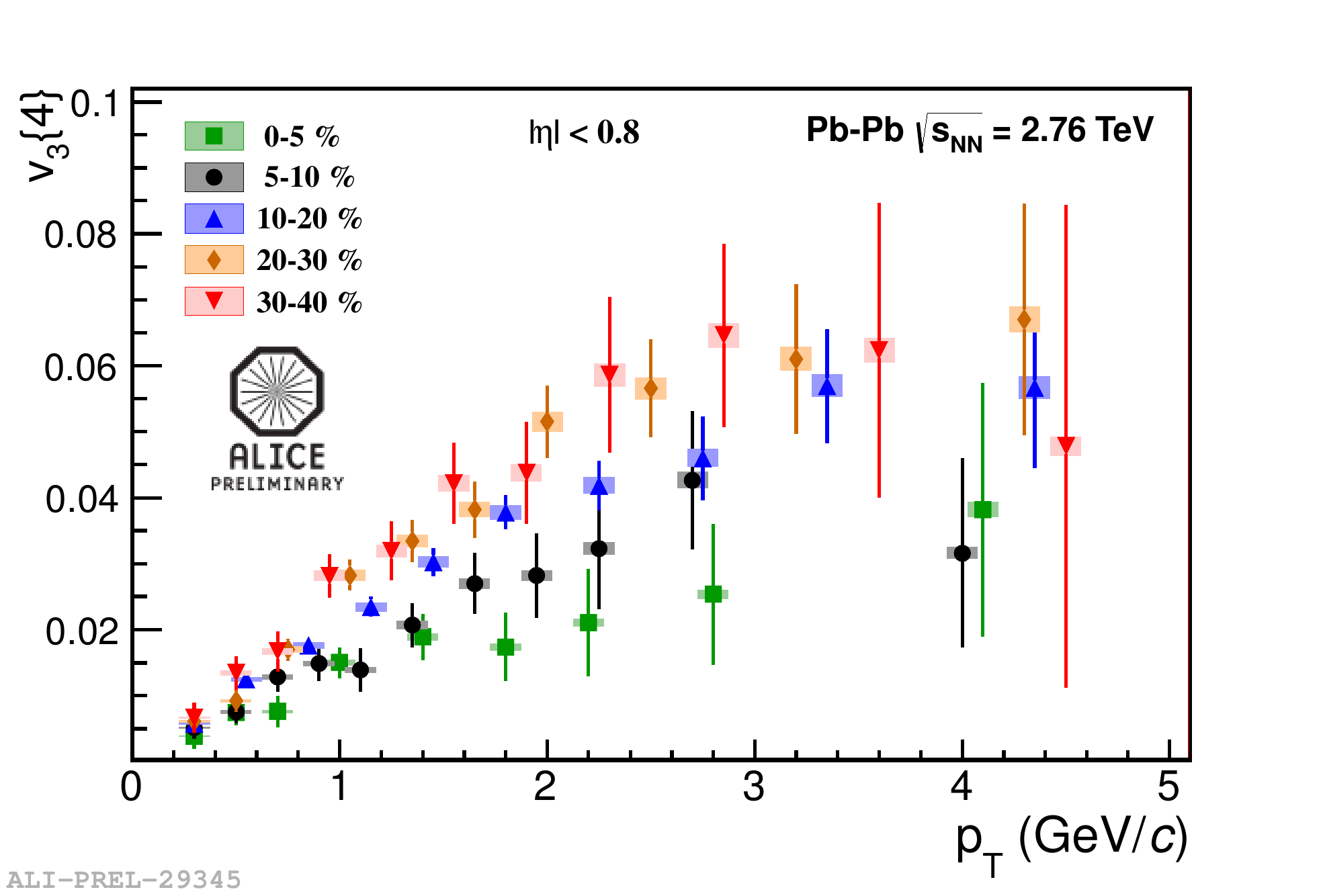}
\includegraphics[width=0.48\textwidth]{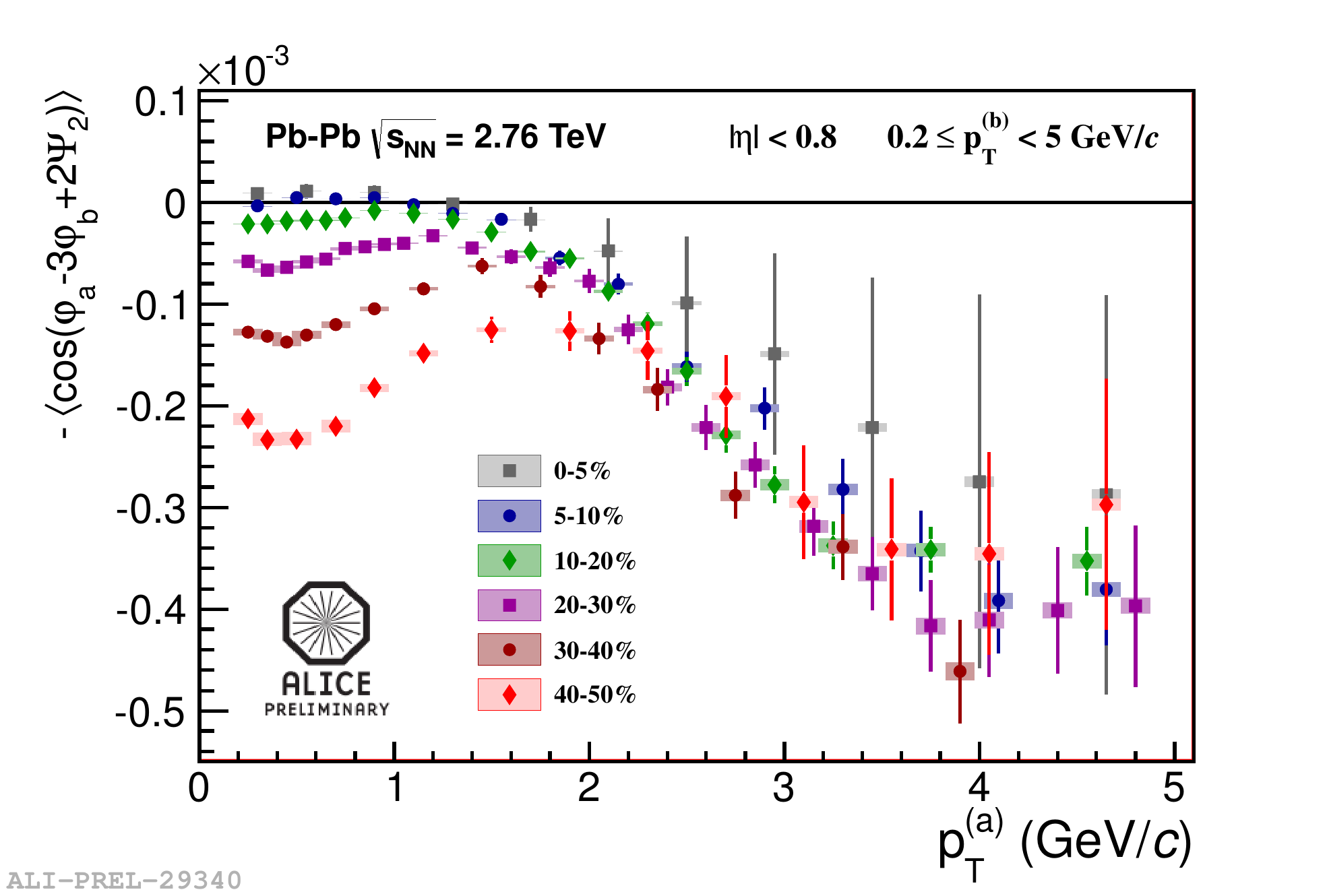}
\end{center}
\vspace*{-4mm}
\caption{$\pt$-differential triangular flow from 4-particle cumulant
  measurement (left panel) and dipole flow from 3-particle cumulants
  (right panel).}
\label{fig:ante}
\end{figure}

%%High multiplicity of the LHC events and good event statistics allow
%precise measurements of multiparticle correlations and thus to study
%the nature of the flow fluctuations in more detail.  
Thanks to the high multiplicity of Pb-Pb collisions at the LHC and to
the large recorded sample of events, precise measurements of
multiparticle correlations and investigation of the nature of flow
fluctuations can be performed in detail.  An example 
is presented in Fig.~\ref{fig:ante} (left panel) where the
differential triangular flow is measured with the 4-particle cumulant
method (for the average flow results as a function of centrality,
see~\cite{ante}).  Figure~\ref{fig:ante} (right panel) presents the
results obtained with 3-particle, mixed harmonic correlator as
a function of the transverse momentum of first harmonic particle.  The
correlation between the so-called dipole flow (originating in dipole
like density fluctuations), the second harmonic, and the third
harmonic event planes has been suggested in~\cite{Teaney:2010vd} . The
measurements, though qualitatively similar to theoretical expectations
at lower transverse momenta, are quite different from predictions at
higher momenta. The origin of a striking similarity in results for
different centralities at $\pt>2$~\gevc~is not clear at the moment.
More discussion of the results obtained with many-particle cumulants
can be found in~\cite{ante}.

%=======================================================================
\section{Charge dependent correlations and the chiral magnetic effect}

\begin{figure}[htbp]
\begin{center}
\includegraphics[width=0.48\textwidth]{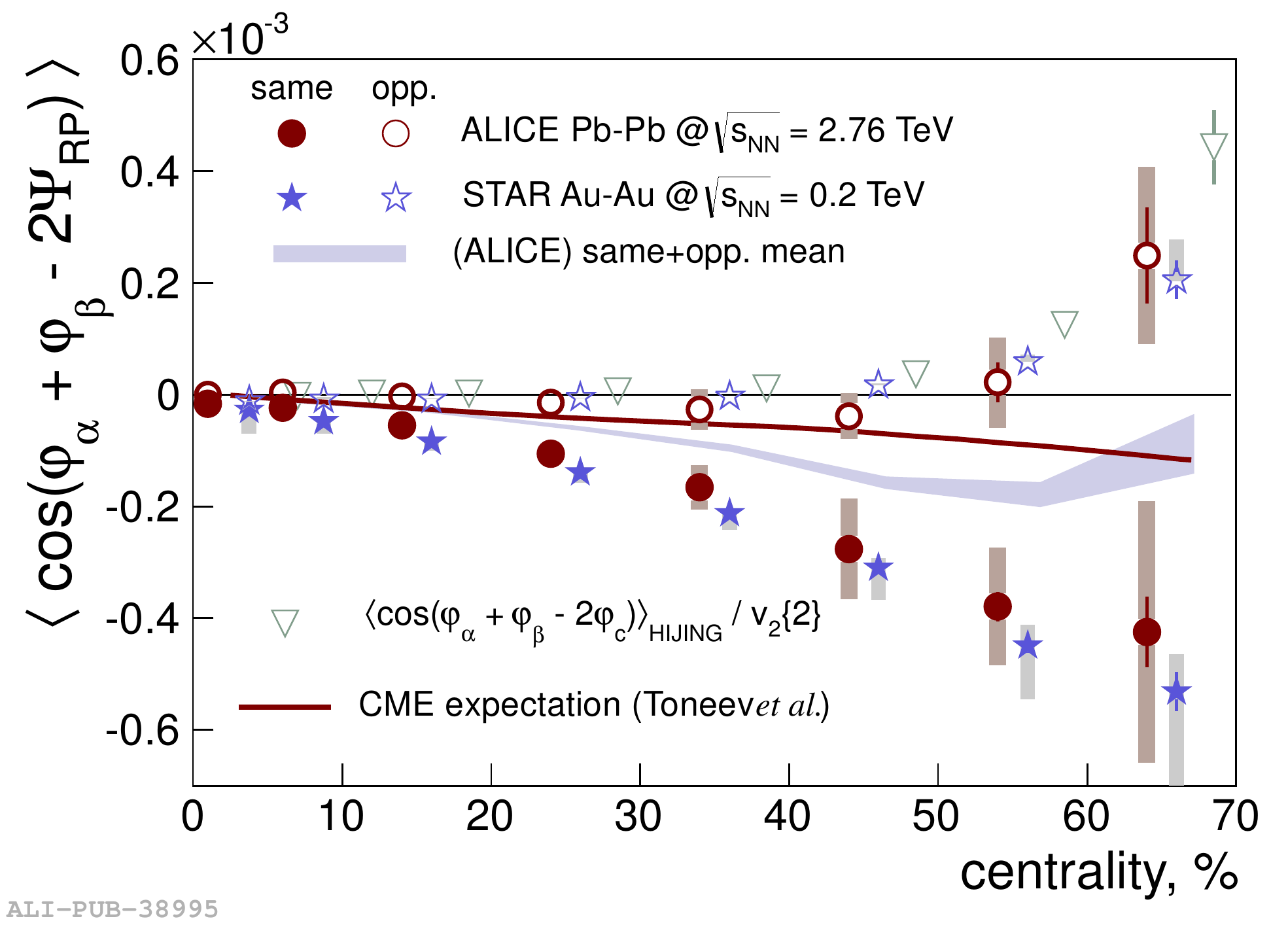}
\includegraphics[width=0.48\textwidth]{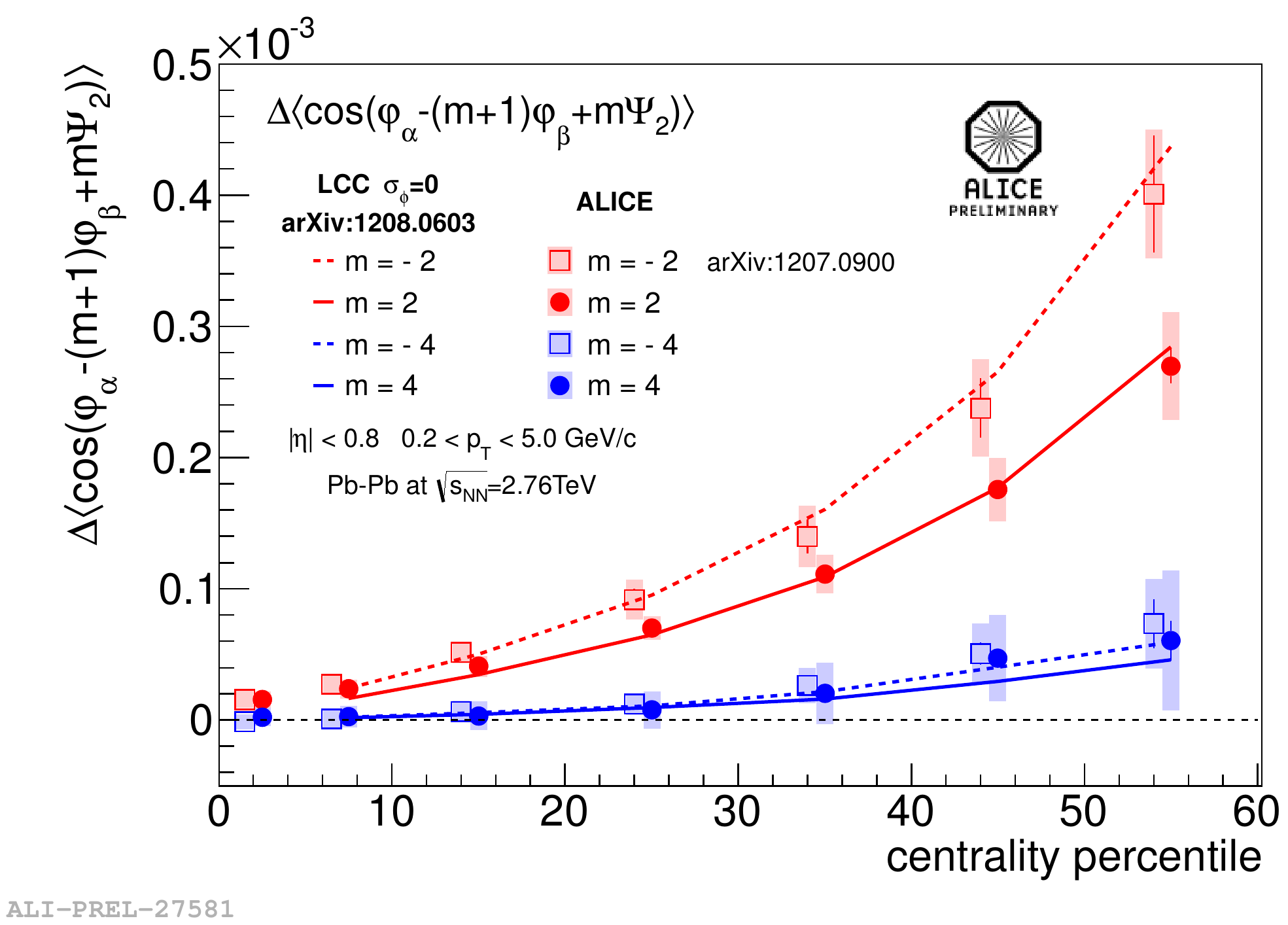}
\end{center}
\vspace*{-4mm}
\caption{(Left panel) Same- and opposite-sign azimuthal
  correlations relative to the second order event plane, and (right
  panel) the charge dependent part of several higher moments of the
  azimuthal correlations compared to the blast wave model calculations.}
\label{fig:hori1}
\end{figure}

Charge dependent correlations, and in particular the charge dependent
correlations relative to the different harmonics event planes are of
great interest as those could reveal the phenomenon of the charge
separation along the magnetic field, usually referred to as the chiral
magnetic effect~\cite{Kharzeev:2004ey}. If observed it would manifest
the local parity violation and directly demonstrate the important role
of instantons and sphalerons, topologically nontrivial solutions of
QCD. The ALICE Collaboration recently measured 2-particle charge
dependent correlations relative to the second order reaction
plane. The results, see the $\pt$-average correlations in the
left panel of Fig.~\ref{fig:hori1},  are qualitatively consistent with
expectations from the chiral magnetic effect. These correlations are
also found to be remarkably similar in strength to those measured at
RHIC.

It is important to note that the observable used for this measurement
is parity even and the results may contain (or even be dominated by)
correlations not related to the parity violation. Although at present
neither event generator can explain these results, the blast wave
model calculations including local charge conservation (at freeze-out)
seem to be able to explain most of the charge dependent part of the
measured correlations~\cite{Schlichting:2010qia}. 
To investigate the effect of local charge conservation in more detail,
the ALICE Collaboration has measured several new correlations, some of
which are presented in Fig.~\ref{fig:hori1} along with results of the
recent blast wave model simulations~\cite{Hori:2012kp}. The agreement
with the model is very good, indicating that the local charge
conservation likely plays an important role in forming such
correlations. Note however that the local charge conservation, at
least as included in these blast wave simulations affects only
opposite charge correlations and has to be combined with some additional
mechanism in order to be consistent with data which exhibits very 
small opposite charge correlations with almost the
entire charge dependent part coming from the same charge correlations.

\begin{figure}[htbp]
\begin{center}
\includegraphics[width=0.8\textwidth]{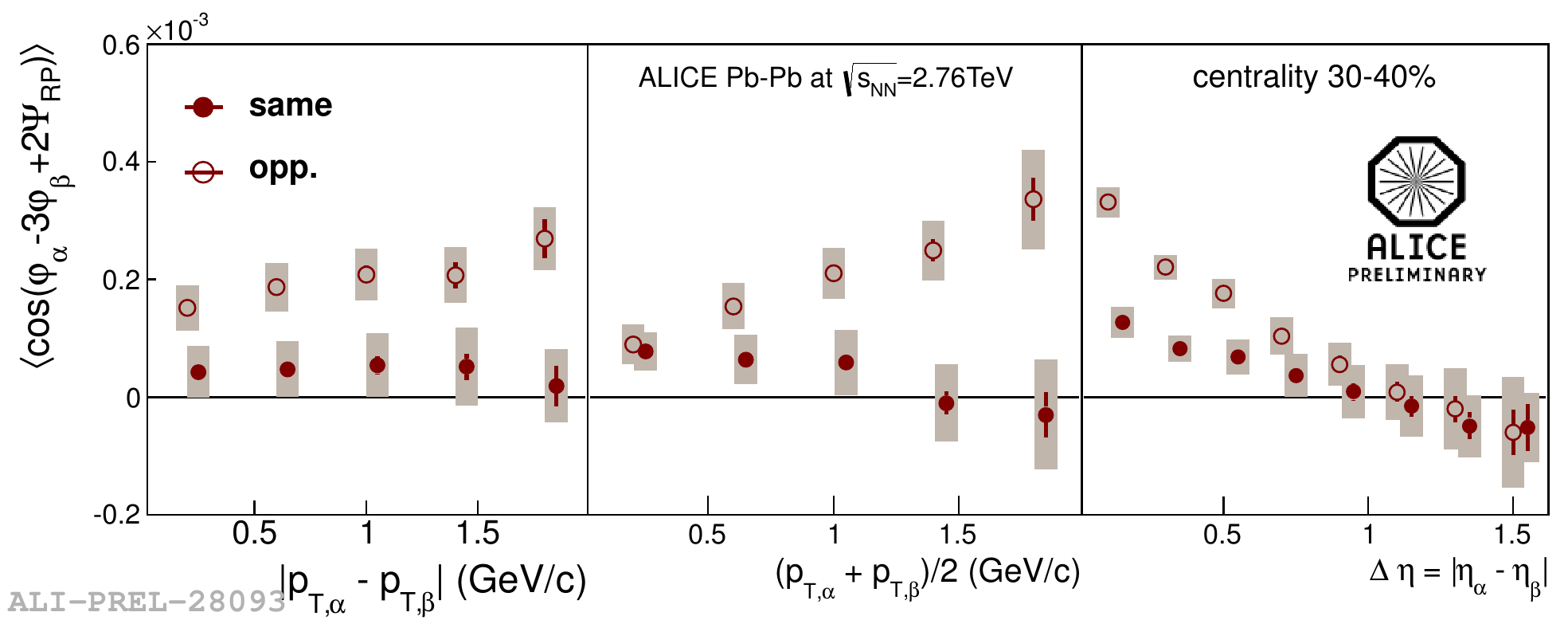}
\end{center}
\vspace*{-4mm}
\caption{Charge dependent correlator
  $\mean{\cos(\phi_\alpha-3\phi_\beta+2\Psi_{\rm RP})}$ as a function of
  the two particle momentum difference, momentum sum, and
  pseudorapidity separation.}
\label{fig:hori2}
\end{figure}

Further insights in the nature of correlation can be obtained with
differential studies, an example of which is shown in
Fig.~\ref{fig:hori2}. Note a possible change in sign in differential
correlations versus $\deta$. It could indicate that the $\deta$
correlations reported in~\cite{Abelev:2012pa} for the 
$\mean{\cos(\phi_\alpha+\phi_\beta-2\Psi_{\rm RP})}$
correlator might not go to zero at large $\deta$ as
originally thought, but only change sign. The
measurements at larger $\deta$ would be very desirable. More details
on ALICE measurements of the charge dependent correlation can be found
in~\cite{hori}.

\begin{figure}[htbp]
\begin{center}
\includegraphics[width=0.48\textwidth]{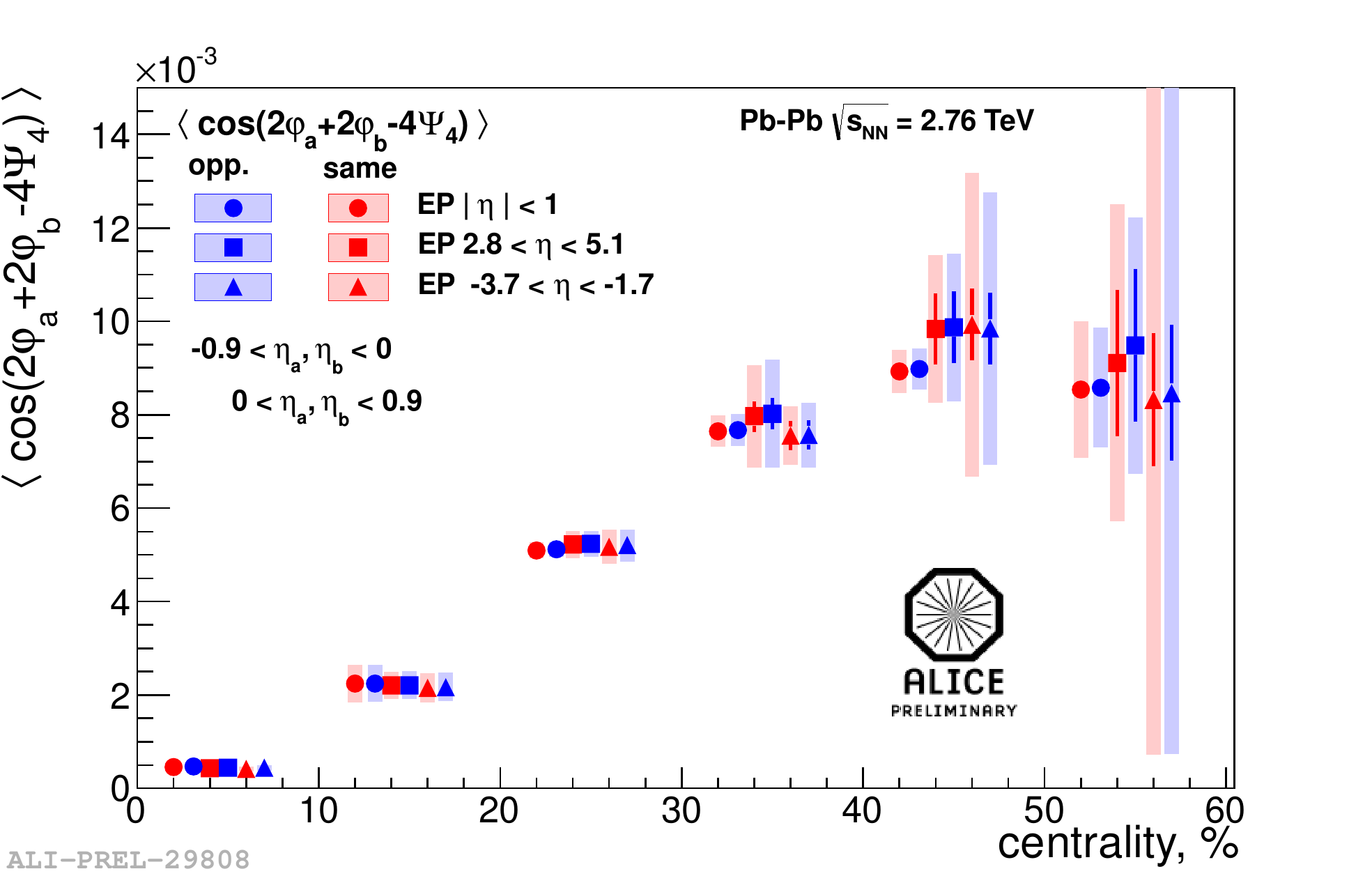}
\includegraphics[width=0.48\textwidth]{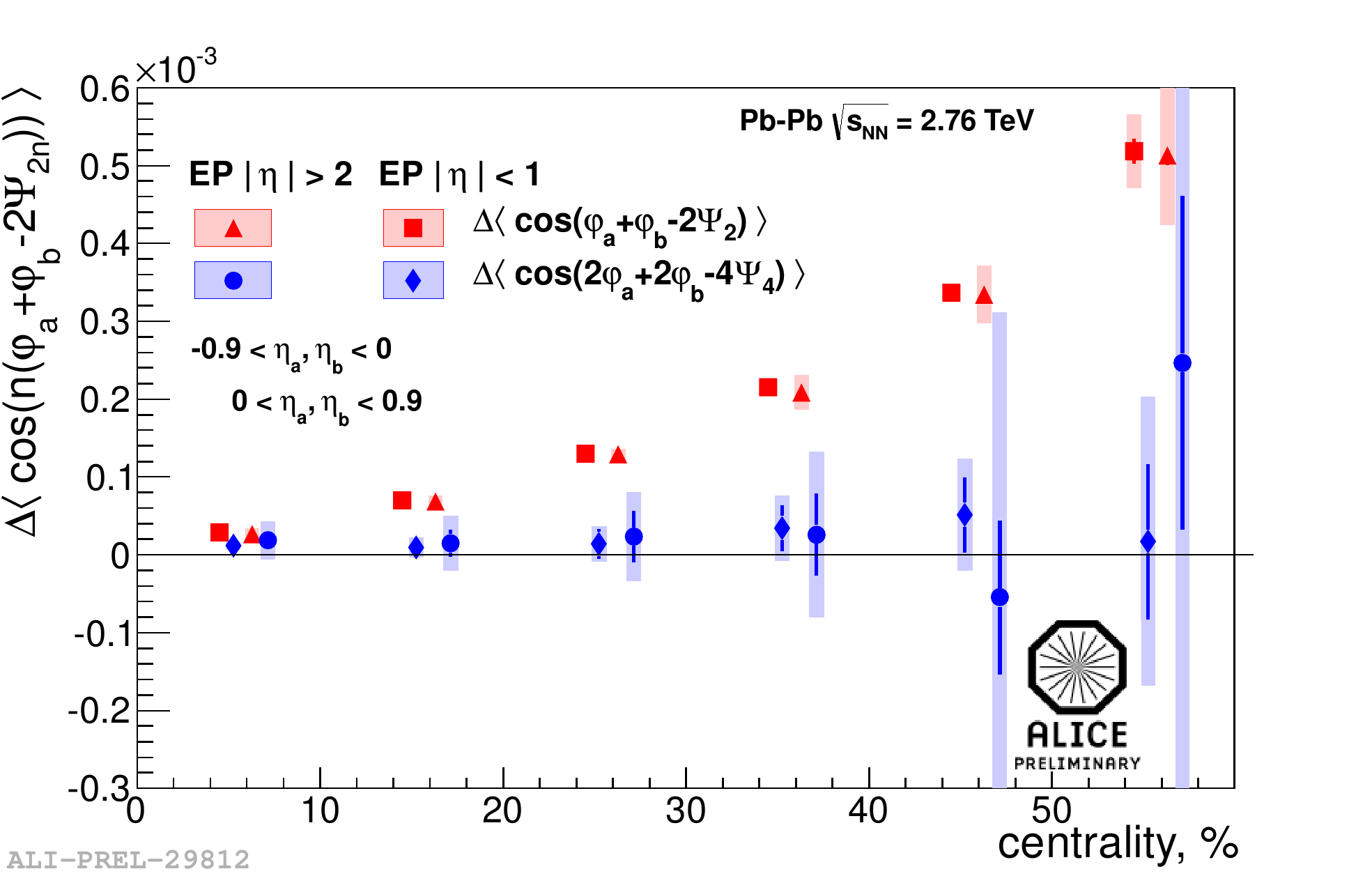}
\end{center}
\vspace*{-4mm}
\caption{(Left panel) Charge dependent correlations relative to the
fourth harmonic event plane as a function of centrality. (Right panel)
Comparison of the charge dependent part of the correlations with respect
to the second and fourth harmonic event planes. }
\label{fig:jm}
\end{figure}

Another measurement that can clarify the origin of the charge
dependent correlations and the role of the local charge conservation
was suggested in~\cite{Voloshin:2011mx}: the correlations measured with
respect to the fourth harmonic even plane should not contain any
contribution from the CME effect but it should include the effect of
local charge conservation. The correlations due to the local charge
conservation in this case are expected to be somewhat smaller in
magnitude as the fourth harmonic flow is not that strong as the elliptic
flow. The results of such measurements are presented in
Fig.~\ref{fig:jm} with charge dependent part shown in the right
panel. The correlations relative to the fourth harmonic event plane
are very weak and suggestive of small contribution from local charge
conservation but the detailed blast wave simulation has to be
performed to draw more definite conclusion from this measurement.

%===================================================================
\section{Event shape engineering}

During the last few years our understanding of the importance of the
initial geometry fluctuations and their role in forming odd harmonic
anisotropic flow and all flow fluctuations has significantly improved.
The results of measurements and simulations using Monte Carlo Glauber
model, suggest that flow fluctuations are very large. This can be used
to select events with unusually large or small eccentricities, the
technique called the event shape
engineering~\cite{Schukraft:2012ah}. This technique allows
unprecedented studies, for example, of the system with very high
density typical of very central collisions, and very large
eccentricities as those in peripheral collisions.  One should be
careful performing the event shape analysis, not to be mistaken with
events dominated by nonflow contribution. It was
suggested~\cite{Schukraft:2012ah} to use event selection (using flow
vector magnitude) based on the information from one momentum window
(subevent) and perform a physical analysis in a different momentum
window (subevent) which is expected to have small correlations via
nonflow to the first one.

\begin{figure}[htbp]
\begin{center}
\includegraphics[width=0.48\textwidth]{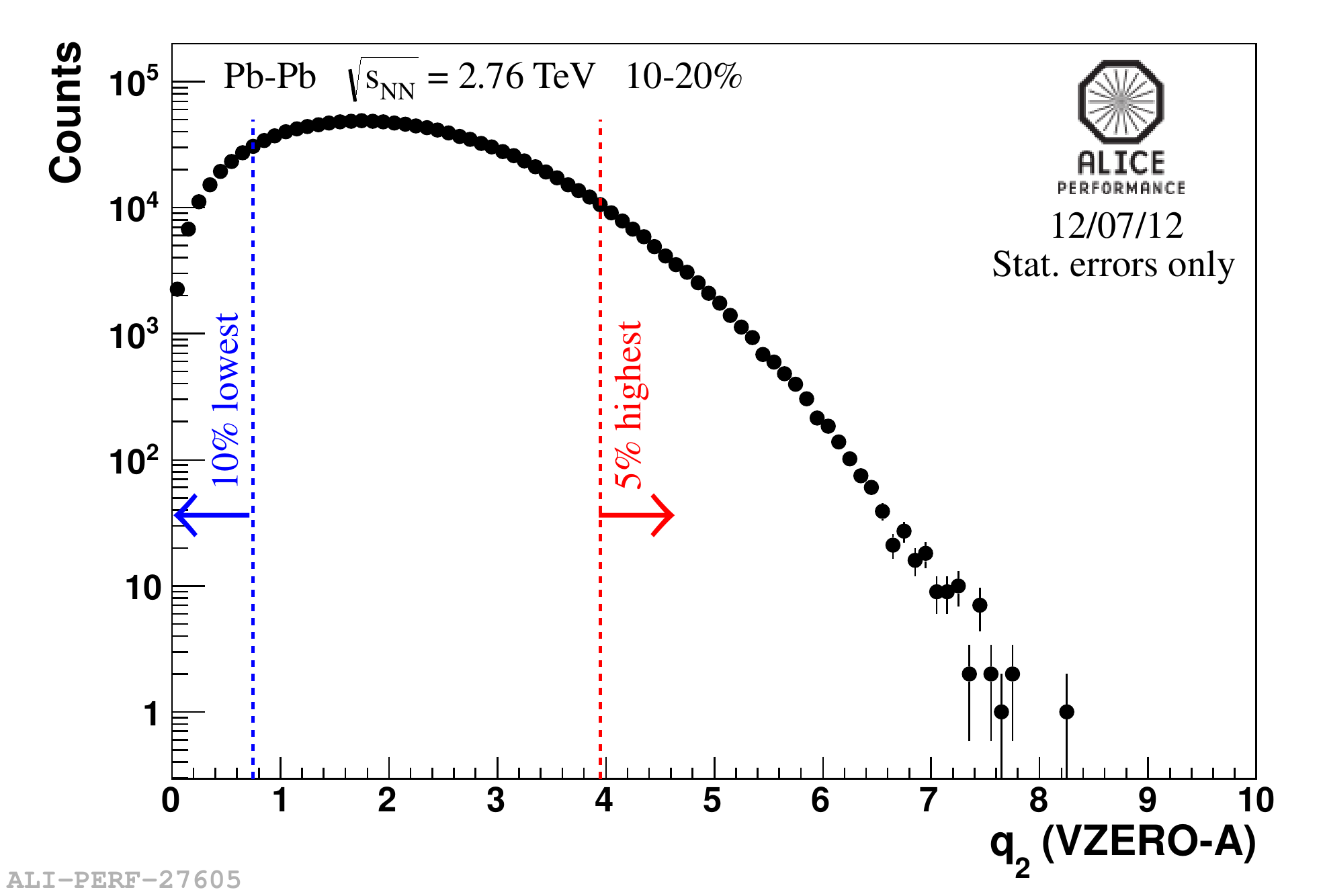}
\includegraphics[width=0.48\textwidth]{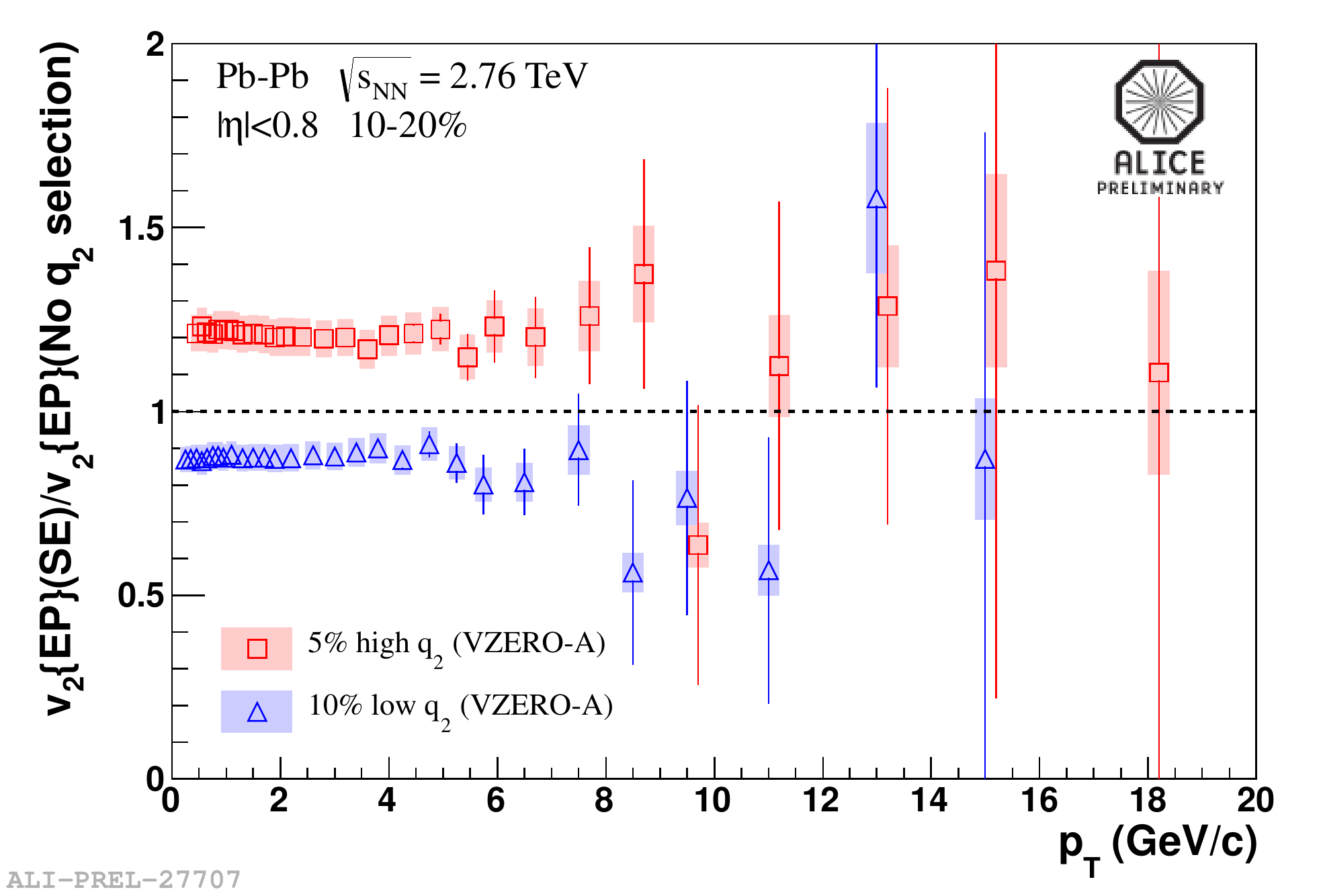}
\end{center}
\vspace*{-4mm}
\caption{(Left panel) Flow vector magnitude distribution and cuts used
  to select events with large and small flow. (Right panel) The ratio
  of flow measured for selected events to that without event shape
  selection as a function of transverse momentum.}
\label{fig:ese}
\end{figure}

In ALICE we use an event selection based on the flow vector magnitude
measured in one of the VZERO detectors and perform the analysis using
the TPC. Those detectors are separated by about two units of
pseudorapidity which greatly suppresses nonflow correlations between
the two~\cite{Abelev:2012di}. The results of this event shape
engineering analysis are presented in Fig.~\ref{fig:ese}, where the
left panel indicates the cuts used to select events with large and
small flow (eccentricity) values.  The right panel shows the results
obtained with the corresponding event selection, in this case the
ratio of elliptic flow as a function of transverse momentum in events
with large and small flow. Approximately constant ratios indicate that
the effect of flow fluctuations on particle production at different
transverse momenta is very similar up to $\pt\approx$10~\gevc, which
is in agreement with earlier conclusions made from comparison of the
two- and four-particle cumulant measurements. More details on this
measurement can be found in~\cite{dobrin}; for the study of $\pt$
spectra modification in events with large or small flow,
see~\cite{milano}.

%==================================================================
\section{Summary}

%ALICE has done great meaurements!!!

The ALICE Collaboration has significantly extended its
flow measurements toward higher transverse momentum and larger
pseudorapidity coverage. It presented numerous results on anisotropic
flow up to $\pt\approx$20~\gevc~ as well as average elliptic and
triangular flow in the region $|\eta|<5$. Based on those measurements
we find that the flow fluctuation effect is very similar at all
pseudorapidities and up to transverse momenta of about 10~\gevc.  It
was found that the difference between proton and pion elliptic flow
extends to about the same transverse momentum, possibly
indicating that this might be the region where jet physics starts to
completely dominate all features of the particle production.
The identified particle flow in the low transverse momentum range is
well described by the hydrodynamical model (including hadronic
afterburner), while in the intermediate $\pt$ region, often associated
with the quark coalescence picture, the NCQ scaling holds at the
10-15\% level. Remarkable is the elliptic flow of $\phi$-meson which
clearly follows the mass splitting in the low transverse momentum
region and follows other meson flow at higher transverse momentum.
The ALICE Collaboration has presented a suit of charge dependent
azimuthal correlations which might significantly clarify their nature,
the role of the local charge conservation and possible contribution to
these correlations from the chiral magnetic effect.
Finally, we have demonstrated that a new promising technique, the event
shape engineering, is fully capable of producing new and important
results.

%============================================================
%\section*{References}

\end{document}